\pdfoutput=1

\documentclass[twocolumn,english]{revtex4}
\usepackage{amssymb,amsmath,graphicx}
\usepackage{color}
\usepackage[T1]{fontenc}
\usepackage{times}
\usepackage{babel}
\begin{document}


\title{
  Graphene disk in a~solenoid magnetic potential: Aharonov-Bohm effect
  without a~two-slit-like setup
}

\author{Adam Rycerz}
\affiliation{Institute for Theoretical Physics,
  Jagiellonian University, \L{}ojasiewicza 11, PL--30348 Krak\'{o}w, Poland}

\author{Dominik Suszalski}
\affiliation{Institute for Theoretical Physics,
  Jagiellonian University, \L{}ojasiewicza 11, PL--30348 Krak\'{o}w, Poland}

\date{April 7, 2020}

\begin{abstract}
  The Aharonov-Bohm effect allows one to demonstrate the physical
  meaningfulness of magnetic vector potential by passing the current
  in zero magnetic field regions. 
  In the standard (a~{\em two-slit-like}) setup a~conducting ring is pierced
  by magnetic flux and the quantum interference for an electron passing
  simultaneously the two ring arms is observed. Here we show, by analyzing the 
  transport via evanescent waves, that the ballistic Corbino disk in graphene
  subjected to a~solenoid magnetic potential may exhibit the conductance
  oscillations of the Aharonov-Bohm kind although the current flows through
  a~single conducting element only. 
\end{abstract}

\maketitle

\section{Introduction}
Quantum transport through the Corbino disk in graphene has been addressed
both theoretically \cite{Ryc09,Ryc10,Kat10,Kha13,Abd17,Jon17,Sus20}
and experimentally \cite{Yan10,Pet14,Kum18,Zen19}
by numerous authors, as the egde-free geometry allows one to probe the
mesoscopic aspects of graphene, such as transport via evanescent waves
\cite{Kat12}, even in nanometer-scale devices. 
At zero magnetic field, conductance of ultraclean ballistic disks as
a~fuction of the carrier concentration \cite{Kum18} shows good agreement
with the basic mode-matching analysis of Ref.\ \cite{Ryc09}. 
At nonzero field, periodic (approximately sinusoidal) magnetoconductance
oscillations were predicted \cite{Ryc10,Kat10} but experimental confirmation
of such a~remarkable quantum-interference phenomenon is missing. 

Theoretical analysis of Ref.\ \cite{Ryc10} employs the rotational symmetry
of the problem, resulting in the total angular momentum conservation
($J_z=\hbar{}j$, with $j=\pm{}1/2,\pm{3/2},\dots$, the angular-momentum
quantum number). 
In the case of an undoped disk of the inner radius $R_1$ and the outer radius
$R_2$, the Landauer-B\"{u}ttiker transmission probabilities \cite{Lan57,But85}
read
\begin{equation}
  \label{tjphd}
  T_j =
  \frac{1}{\cosh^2\left[\ln(R_2/R_1)\left(j+\Phi_d/\Phi_0\right)\right]},
\end{equation}
where $\Phi_d=\pi{}(R_2^2-R_1^2)B$ is the flux piercing the disk with
a~uniform magnetic field $B$, and $\Phi_0=2(h/e)\ln(R_2/R_1)$ defines the
conductance-oscillation period. 
Further analysis shows that the formula equivalent to given by Eq.\
(\ref{tjphd}) can also be derived if the carrier concentration
(hereinafter quantified by the Fermi energy $E_F$, with $E_F=0$ corresponding
to the charge-neutrality point) is adjusted to any
Landau level, $E_n = \mbox{sgn}(n){}v_F\sqrt{2|n|eB}$, with
$n=0,\pm{}1,\pm{2},\dots$, and $v_F\approx{}10^{6}\,$m/s
being the energy-independent Fermi velocity in graphene. 

Away from Landau levels the transmission is strongly suppressed
\cite{Ryc10,Sus20}. For instance, in the vicinity of the charge-neutrality
point ($n=0$) magnetoconductance oscillations may be observed
in the magnetic field range limited by
\begin{equation}
\label{phbound}
|\Phi_d|\lesssim{}\frac{2h}{e}\ln\left(\frac{1}{k_F{}R_1}\right)
= \frac{h}{e}\ln\left(\frac{1}{\pi{}|n_C|R_1^2}\right), 
\end{equation}
where we have further defined $k_F=|E_F|/(\hbar{}v_F)$ and outermost right
equality follows from the relation beetween the Fermi wavenumber and the
carrier concentration ($n_C$), namely $k_F=\sqrt{\pi|n_C|}$,
including the fourfould (spin and valley) degeneracy of each
quasiparticle level.
On the other hand, the current flow through the system leads to carrier density
fluctuations of the order of $\delta{}n_C\sim{}1/(\pi{}{R_2}^2)$, even in the
absence of the charge inhomogeneity usually appearing due to the electron-hole
puddle formation at low densities \cite{Dea10}.
Taking the above as the lower bound to $|n_C|$ in Eq.\ (\ref{phbound}),
one immediately obtains $|\Phi_d|\lesssim{}\Phi_0$, suggesting it may be
difficult (or even impossible) to observe the magnetoconductance oscillations
in the linear-response regime. A~proposal to overcome this diffuculty by
going beyond the linear-response regime was put forward \cite{Rut15}.

\begin{figure}[!b]
  \includegraphics[width=0.7\linewidth]{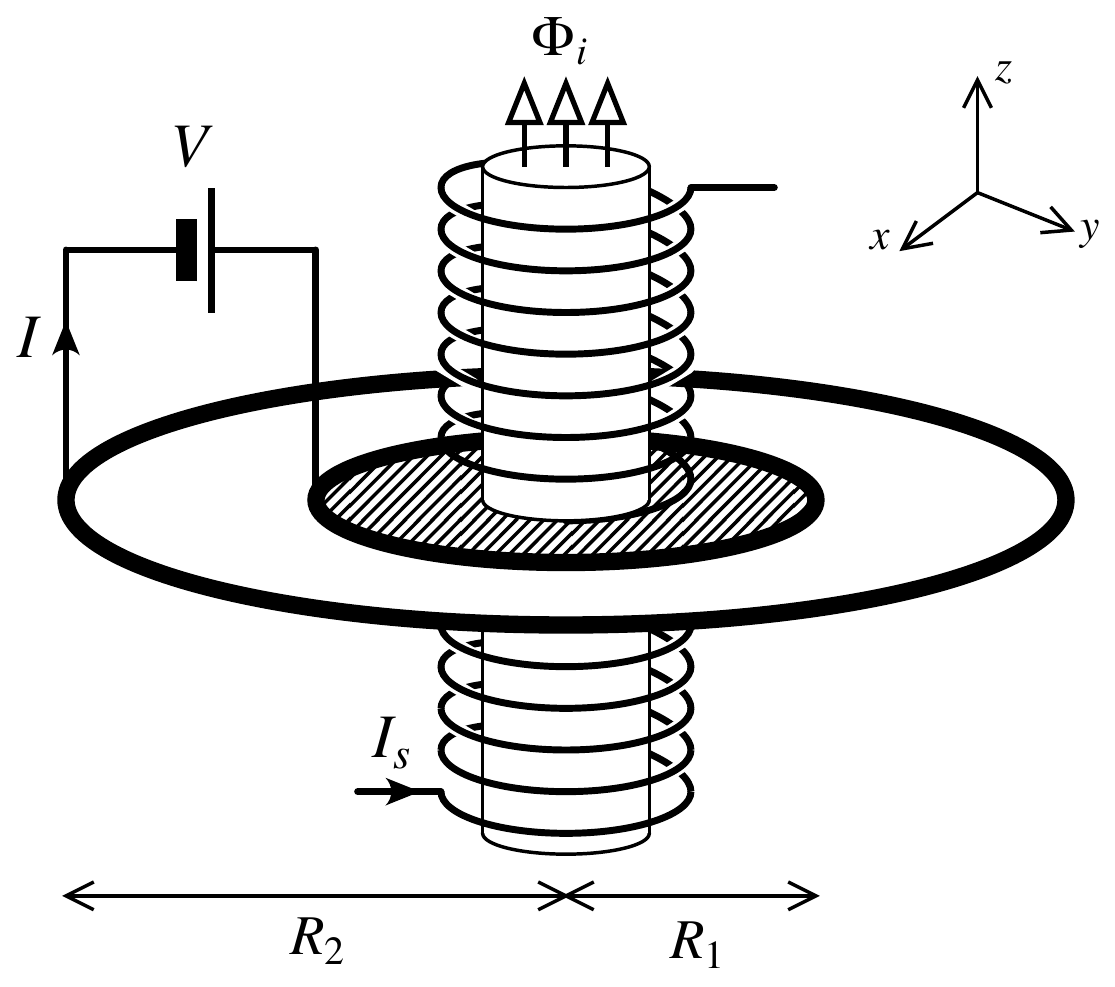}
\caption{ \label{phidisk}
  Schematic of the Corbino disk in graphene of the inner radius $R_1$ and
  the outer radius $R_2$, contacted by two electrodes (tick black circles).
  A~voltage source ($V$) drives the current ($I$) through the disk.
  A~separate gate electrode (not shown) allows the carrier concentration
  in the disk to be tuned around the neutrality point.
  A~long solenoid, carrying the current $I_s$, generates the flux
  $\Phi_i$ piercing the inner disk area.
  The coordinate system is also shown. 
}
\end{figure}

A~separate issue concerns the role of electron-electron interactions,
which is usually marginal when discussing ballistic systems in monolayer
graphene \cite{Sie11,Hwa12},
in agreement with fundamental considerations \cite{Mar97,Dus13}, 
but may lead to Wigner crystallization or the appearance of fractional
quantum Hall phases, 
in case the bulk density of states is strongly modified due to the Landau
quantization \cite{Kum18,Aba10,Lee12}. 


Generally speaking, uniform magnetic fields, although being most feasible to
generate at micrometer scale, do not seem to provide a~realistic opportunity
to observe magnetoconductance oscillations in graphene-based Corbino disks.
Therefore, it is worth to consider another field arrangemets, in which phase
effects may overrule orbital effects (such as the Landau level formation).
In this paper, we focus on the case of the disk which inner area is pierced
by a~long solenoid (see Fig.~\ref{phidisk}), generating the flux $\Phi_i$.
Earlier, it was shown by Katsnelson \cite{Kat10,magfoo} that for zero doping
($E_F=0$) the transmission probabilities are given by Eq.\ (\ref{tjphd})
after substituting
\begin{equation}
  \Phi_d\equiv{}\Phi_i\ \ \ \ \ 
  \text{and}\ \ \ \ \ 
  \Phi_0\equiv{}\Phi_{\rm AB}, 
\end{equation}
with $\Phi_{\rm AB}=h/e$ being the familiar Aharonov-Bohm flux quantum
\cite{Naz09}.
However, the analysis of such a~system away from the charge-neutrality point
($E_F\neq{}0$) is missing. 

The remaining part of the paper is organized as follows.
In Sec.\ \ref{solut} we present the results of the mode-matching analysis
for the system of Fig.~\ref{phidisk} at arbitrary doping and flux.
Next, in Sec.\ \ref{resul}, the numerical discussion of the conductance
oscillations is provided.
The effect of electrostatic field breaking cylindrical symmetry of the
problem is considered in Sec.\ \ref{nosymm}. 
The conclusions are given in Sec.\ \ref{conclu}.

\section{Solution for arbitrary doping and flux}
\label{solut}

Our analysis starts from the Dirac Hamiltonian in a single valley ($K$),
which is given by
\begin{equation}
  \label{hamdisk}
  H = v_F{}\left(\mbox{\boldmath$p$}+e\mbox{\boldmath$A$}\right)\cdot
  \mbox{\boldmath$\sigma$} + U(r),
\end{equation}
where $\mbox{\boldmath$p$}=-i\hbar{}(\partial_x,\partial_y)$ is the in-plane
momentum operator, the electron charge is $-e$, 
the magnetic vector potential of a~solenoid is written in the
symmetric gauge \cite{Fey10}
\begin{equation}
  \label{vecsol}
  \mbox{\boldmath$A$} = \left(A_x,A_y\right) =
  \frac{\Phi_i}{2\pi{}}\left(-\frac{y}{r^2}, \frac{x}{r^2}\right),
\end{equation}
and $\mbox{\boldmath$\sigma$}=(\sigma_x,\sigma_y)$ with
$\sigma_x$ and $\sigma_y$ being the Pauli matrices.
We further suppose that the electrostatic potential energy $U(r)$
depends only on $r=\sqrt{x^2+y^2}$; namely, we put $U(r)=0$ in the disk
area ($R_1<r<R_2$), or $U(r)=U_\infty$ otherwise. 
Since Hamiltonian (\ref{hamdisk}) commutes with the total angular momentum
operator, $J_z=-i\hbar{}\partial_\varphi+\hbar{}\sigma_z/2$, the energy
eigenfunctions can be chosen as eigenstates of $J_z$ 
\begin{equation}
\label{psirphi}
  \psi_j(r,\varphi) = e^{i(j-1/2)\varphi}\left(
  \begin{array}{c}
    \chi_{j,A}(r) \\
    \chi_{j,B}(r)e^{i\varphi} \\
  \end{array}
  \right), 
\end{equation}
where $j$ is a~half-odd integer, two spinor components ($A$, $B$)
correspond to the sublattice degree of freedom, and we have introduced
the polar coordinates $(r,\varphi)$. 
The Dirac equation now can be written as $H_j(r)\chi_j(r)=E\chi_j(r)$, 
where $\chi_j(r)=[\chi_{j,A}(r),\chi_{j,B}(r)]^T$, and
\begin{multline}
\label{hamjr}
  H_j(r)=-i\hbar{v_F}\sigma_x\partial_r+U(r) \\
  +\hbar{v_F}\sigma_y\left(\begin{array}{cc}
     \frac{\displaystyle{}j-1/2}{\displaystyle{}r}
     +\frac{\displaystyle{}e\Phi_i}{\displaystyle{}hr} & 0 \\
     0 & \frac{\displaystyle{}j+1/2}{\displaystyle{}r}
     +\frac{\displaystyle{}e\Phi_i}{\displaystyle{}hr} \\
  \end{array}\right).
\end{multline}

For a~piecewise-constant potential energy $U(r)$ and the electron-doping
case, $E>U(r)$, the eigenfunctions of $H_j(r)$ (\ref{hamjr}) for the
incoming (i.e., propagating from $r=0$) and outgoing (propagating from
$r=\infty$) waves are given, up to the normalization, by
\begin{equation}
  \chi_j^{\rm in} \!= \left(
  \begin{array}{c}
    H_{\nu(j)-1/2}^{(2)}(kr) \\
    iH_{\nu(j)+1/2}^{(2)}(kr) \\
  \end{array}
  \right),
  \ 
  \chi_j^{\rm out} \!= \left(
  \begin{array}{c}
    H_{\nu(j)-1/2}^{(1)}(kr) \\
    iH_{\nu(j)+1/2}^{(1)}(kr) \\
  \end{array}
  \right),
  \end{equation}
where
\begin{equation}
\label{nujdef}
  \nu(j) = j + \Phi_i/\Phi_{\rm AB},
\end{equation}
$H_\nu^{(1,2)}(\rho)$ is the Hankel
function of the (first, second) kind, and $k=|E-U(r)|/(\hbar{}v_F)$. 
The solution for the disk area can be represented as
\begin{equation}
  \label{chijdisk}
  \chi_j^{(d)}=A_j\chi_j^{\rm in}(k_Fr)+B_j\chi_j^{\rm out}(k_Fr), \ \ \ 
  R_1\!<\!r\!<\!R_2,
\end{equation}
with $A_j$ and $B_j$ being arbitrary constants, and the Fermi wavenumber 
$k_F=|E|/(\hbar{}v_F)$. 
For the hole doping case, $E<U(r)$, the wavefunctions are replaced by
$\tilde{\chi}_j^{\rm in(out)}=\left[\chi_j^{\rm in(out)}\right]^\star$, where we
use the relation $H_\nu^{(2)}=\big[H_\nu^{(1)}\big]^\star$. 

The heavily-doped graphene leads are modeled here by taking the limit of
$U(r)=U_\infty\rightarrow{}\pm\infty$ for $r<R_1$ or $r>R_2$.
The corresponding wavefunctions can be simplified to
\begin{align}
  \chi_j^{(1)} &=
    \frac{e^{\pm{}ik_\infty}}{\sqrt{r}}
    \left(\begin{array}{c} 1 \\ 1 \\ \end{array}\right)
    + r_j
    \frac{e^{\mp{}ik_\infty}}{\sqrt{r}}
    \left(\begin{array}{c} 1 \\ -1 \\ \end{array}\right),
    \  &r<R_1,
  \label{chij1} \\
  \chi_j^{(2)} &= t_j
    \frac{e^{\pm{}ik_\infty}}{\sqrt{r}}
    \left(\begin{array}{c} 1 \\ 1 \\ \end{array}\right),
    \  &r>R_2, \label{chij2} 
\end{align}
where we have introduced the reflection (transmission) amplitudes $r_j$
($t_j$) and $k_\infty=|E-U_\infty|/(\hbar{}v_F)\rightarrow{}\infty$.

Solving the mode-matching conditions, $\chi_j^{(1)}(R_1)=\chi_j^{(d)}(R_1)$
and $\chi_j^{(d)}(R_2)=\chi_j^{(2)}(R_2)$, 
we find the transmission probability for $j$-th mode
\begin{equation}
\label{tjphi}
  T_{j} = |t_j|^2 = \frac{16}{\pi^2{}k^2{}R_1{}R_2}\,
  \frac{1}{\left[\mathfrak{D}_{\nu(j)}^{(+)}\right]^2
    + \left[\mathfrak{D}_{\nu(j)}^{(-)}\right]^2},
\end{equation}
where $\nu(j)$ is given by Eq.\ (\ref{nujdef}) and  
\begin{multline}
\label{ddnupm}
  \mathfrak{D}_{\nu}^{(\pm)} = \mbox{Im}\left[
    H_{\nu-1/2}^{(1)}(kR_1)H_{\nu\mp{}1/2}^{(2)}(kR_2)\right. \\
    \pm \left.H_{\nu+1/2}^{(1)}(kR_1)H_{\nu\pm{}1/2}^{(2)}(kR_2)
    \right]. 
\end{multline}


\section{Results and discussion}
\label{resul}

The linear-response conductance is calculated according to the
Landauer-B\"{u}ttiker formula \cite{Lan57,But85}
\begin{equation}
\label{gsumtj}
  G =\frac{I}{V}= g_0\sum_{j=\pm{}1/2,\pm{}3/2,\dots}T_j, 
\end{equation}
where the conductance quantum $g_0=4e^2/h$, with the factor $4$ accounting
for spin and valley degeneracy, and the summation over modes is performed
numerically up to the machine round-off errors \cite{numfoo}.

\begin{figure}[!t]
  \includegraphics[width=\linewidth]{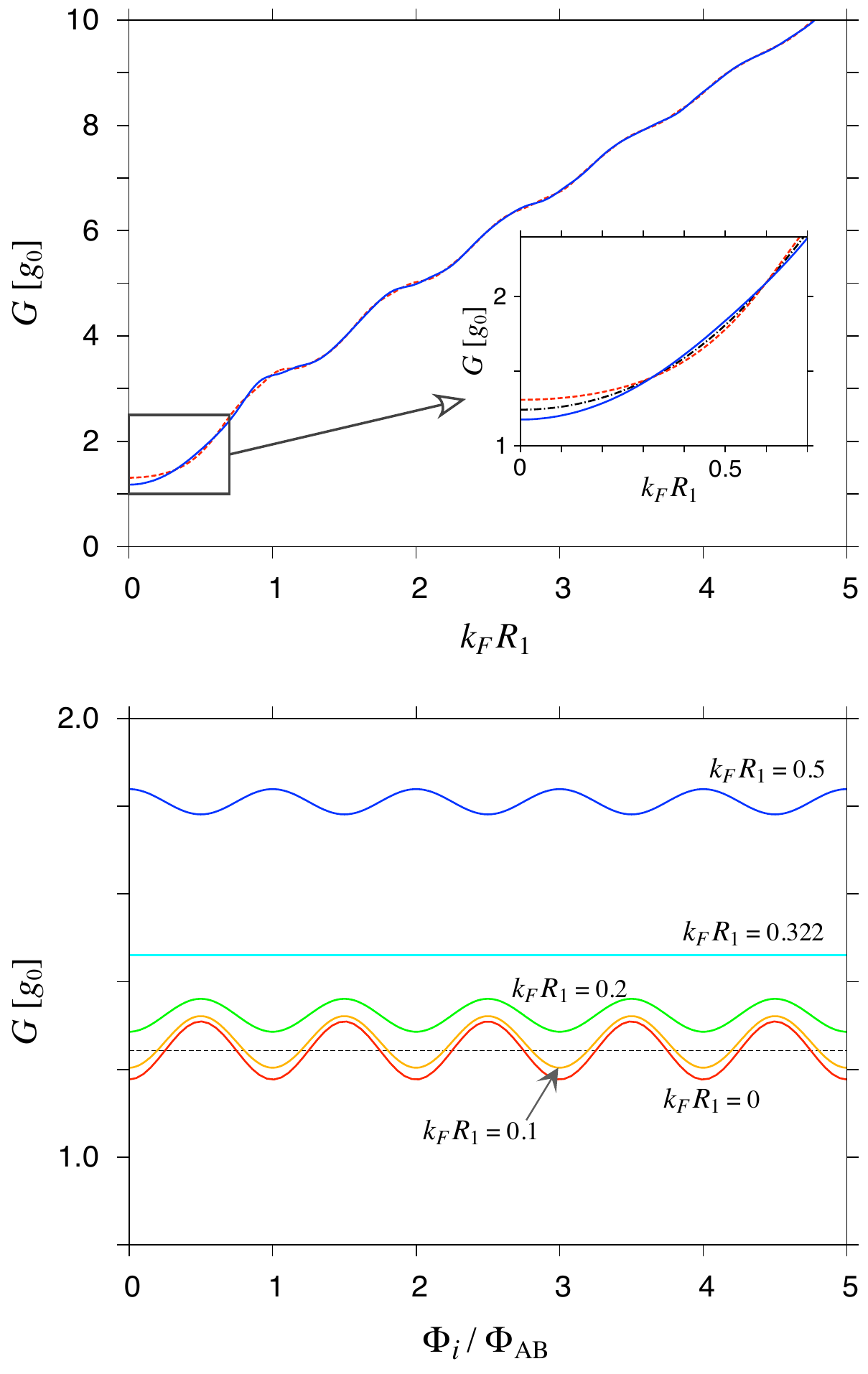}
\caption{ \label{gr5kphi}
  Conductance as a~function of the doping (top) and the flux piercing
  the inner disk area (bottom) for the radii ratio $R_2/R_1=5$.
  Top: Different lines correspond to $\Phi_i=0$ (blue solid line)
  and $\Phi_i=\Phi_{AB}/2$ (red dashed line). Inset presents a~zoom-in,
  for low dopings, with an additional black dash-dot line depicting
  the conductance averaged over $\Phi_i$.
  Bottom: The doping is varied from $k_FR_1=0$ to $k_FR_1=0.5$ and specified
  for each solid line on the plot. Dashed line
  marks the pseudodiffusive conductance $G_{\rm diff}=2g_0/\ln(R_2/R_1)$,
  with $g_0=4e^2/h$. 
}
\end{figure}

\begin{figure}[!t]
  \includegraphics[width=\linewidth]{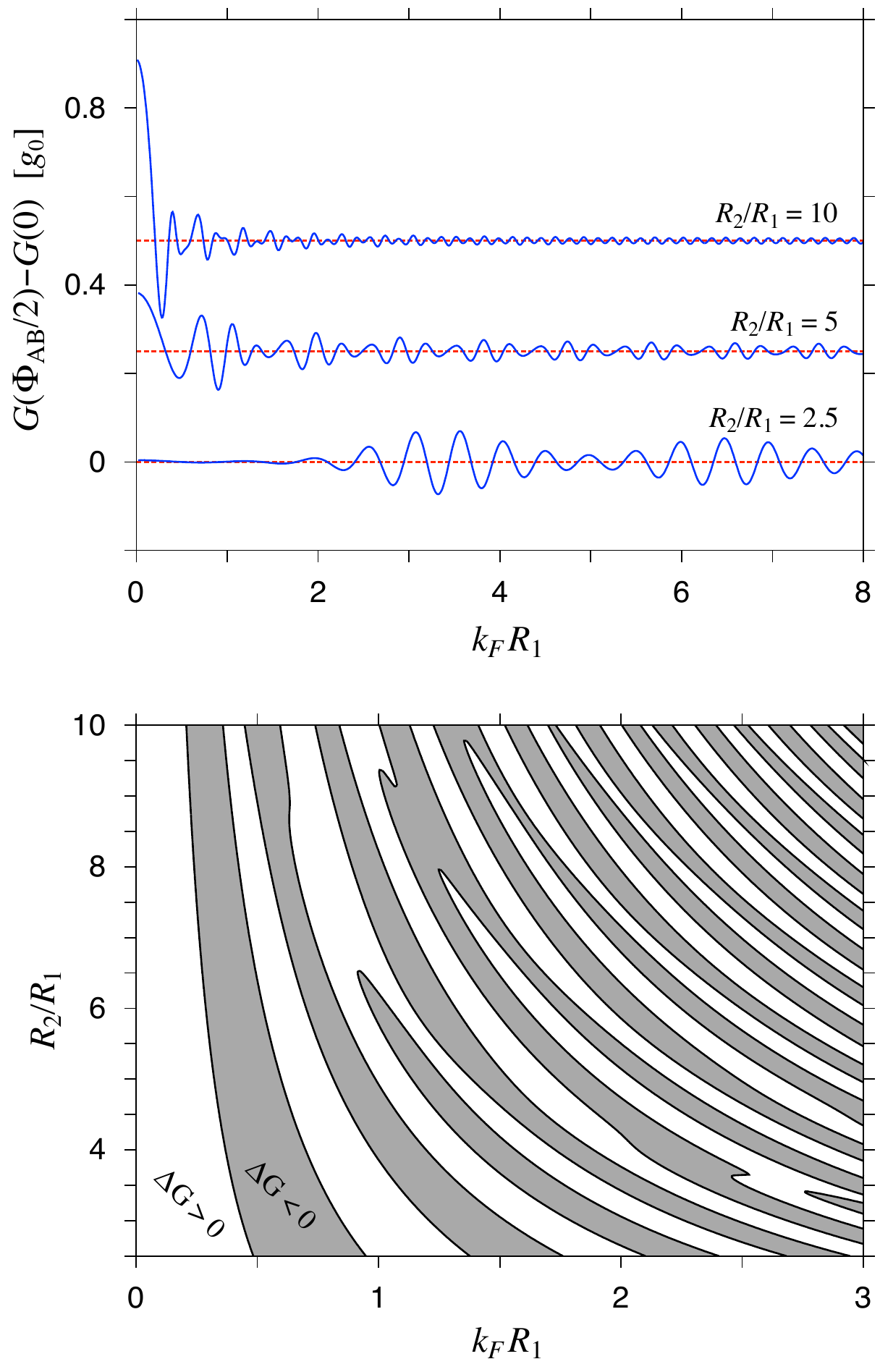}
\caption{ \label{delgnodes}
  Top: Magnitude of the conductance oscillations,
  $\Delta{}G=G(\Phi_{\rm AB}/2)-G(0)$, displayed as a~function of the doping
  for selected values of the radii ratio (specified for each line).
  Notice that the datasets for $R_2/R_1=5$ and $10$ (blue solid lines)
  are subjected to vertical shifts of $0.25$ and $0.5g_0$ (respectively).
  Red dashed line shows the actual $\Delta{}G=0$ for each case.
  Bottom: Nodal lines (black solid) of $\Delta{}G$ as a~function of
  the doping and the radii ratio, separating the areas with $\Delta{}G>0$
  (white) and $\Delta{}G<0$ (shadow). 
}
\end{figure}

Our numerical results are presented in Figs.\ \ref{gr5kphi} and
\ref{delgnodes}. 

The asymptotic properties of the Hankel functions \cite{Nem17} in
Eq.\ (\ref{tjphi}) lead to $T_j\approx{}1$ for $k_FR_1-\nu(j)\gg{}1$, 
with $\nu(j)$ given by Eq.\ (\ref{nujdef}), or to 
$T_j\approx{}0$ for $\nu(j)-k_FR_1\gg{}1$. In turn, the conductivity
can by approximated as $G\approx{}2g_0k_FR_1$ for $k_FR_1\gg{}1$ and
$R_2\gg{}R_1$ (see top panel in Fig.\ \ref{gr5kphi}), with an excess value
of $\sim{}g_0R_1/R_2$ (up to the order of magnitude) representing
the contribution from evanescent waves. 

Furthermore, the structure of Eqs.\ (\ref{nujdef}), (\ref{tjphi}),
(\ref{ddnupm}), and (\ref{gsumtj})
results in perfectly periodic functional dependence of $G(\Phi)$,
with a~period $\Phi_{\rm AB}$, at arbitrary doping (see Fig.\ \ref{gr5kphi}). 
Quite surprisingly, the magnitude of the conductance oscillations
\begin{equation}
\label{delgdef}
  \Delta{}G=G(\Phi_{\rm AB}/2)-G(0),
\end{equation}
takes relatively large absolute values (namely, $|\Delta{}G|>0.1\,g_0$)
not only in small vicinity of the charge neutrality-point, but also
at higher dopings (see Fig.\ \ref{delgnodes}), signaling the importance
of transport via evanescent waves again.
[Notice that the difference between $G(\Phi_{\rm AB}/2)$ and $G(0)$, defining
$\Delta{}G$ via Eq.\ (\ref{delgdef}), is governed by only a~few modes for
which $k_FR_1\approx{}\nu(j)$ and thus $T_j$-s are neither $\,\approx{}0$
nor $\,\approx{}1$.] A~systematic growth of $\Delta{}G$ with $R_2/R_1$
is visible for $k_F\rightarrow{}0$ (with $\Delta{}G\approx{}g_0$ for
$R_2\gg{}R_1$), in consistency with earlier predictions of Refs.\
\cite{Ryc10,Kat10} for the uniform magnetic field case.

For each radii ratio, one can find a~unique series of discrete doping
values for which $\Delta{}G=0$, resulting in $G(\Phi_i)=\text{const}$.
For instance, if $R_2/R_2=5$, the first five nodes of $\Delta{}G$
correspond to
\begin{equation}
\label{degnods}
  \left(k_F{}R_1\right)_{\Delta{}G=0} = 0.322,\ 0.598,\ 0.814,\ 0.987,\ 1.137. 
\end{equation}
Below the first nodal value (i.e., $|k_FR_1|<0.322$), we have $\Delta{}G>0$
[or, equivalently, $G(\Phi_{\rm AB})>G(0)$, see Eq.\ (\ref{delgdef})]; 
then, the sign of $\Delta{}G$ alternates with growing $k_FR_1$, as indicated
in the bottom panel in Fig.\ \ref{delgnodes}. 

It is also visible in Fig.\ \ref{delgnodes} that the pattern of nodal lines
is rather irregular, as one could expect since $\Delta{}G$ can be regarded
as the rational expression containing Bessel function. 
Typical separation between the first nodes of $\Delta{}G$ in Eq.\
(\ref{degnods}) can (roughly) be approximated as
$\Delta{}k_FR_1\approx{}0.3$, which corresponds, for the  physical size of
$R_1=50\,$nm, to the energy interval of
$\Delta{}E_F/k_B\approx{}40\,$K (with the Boltzmann constant $k_B$).
In turn, the conductance oscillations should be observable
in comparable or higher temperatures then the standard
Aharonov-Bohm effect in graphene rings \cite{Rus08,Sta09}.

%
%

\section{Conductance oscillations in the absence of cylindrical symmetry}
\label{nosymm}

\begin{figure}[!t]
  \includegraphics[width=\linewidth]{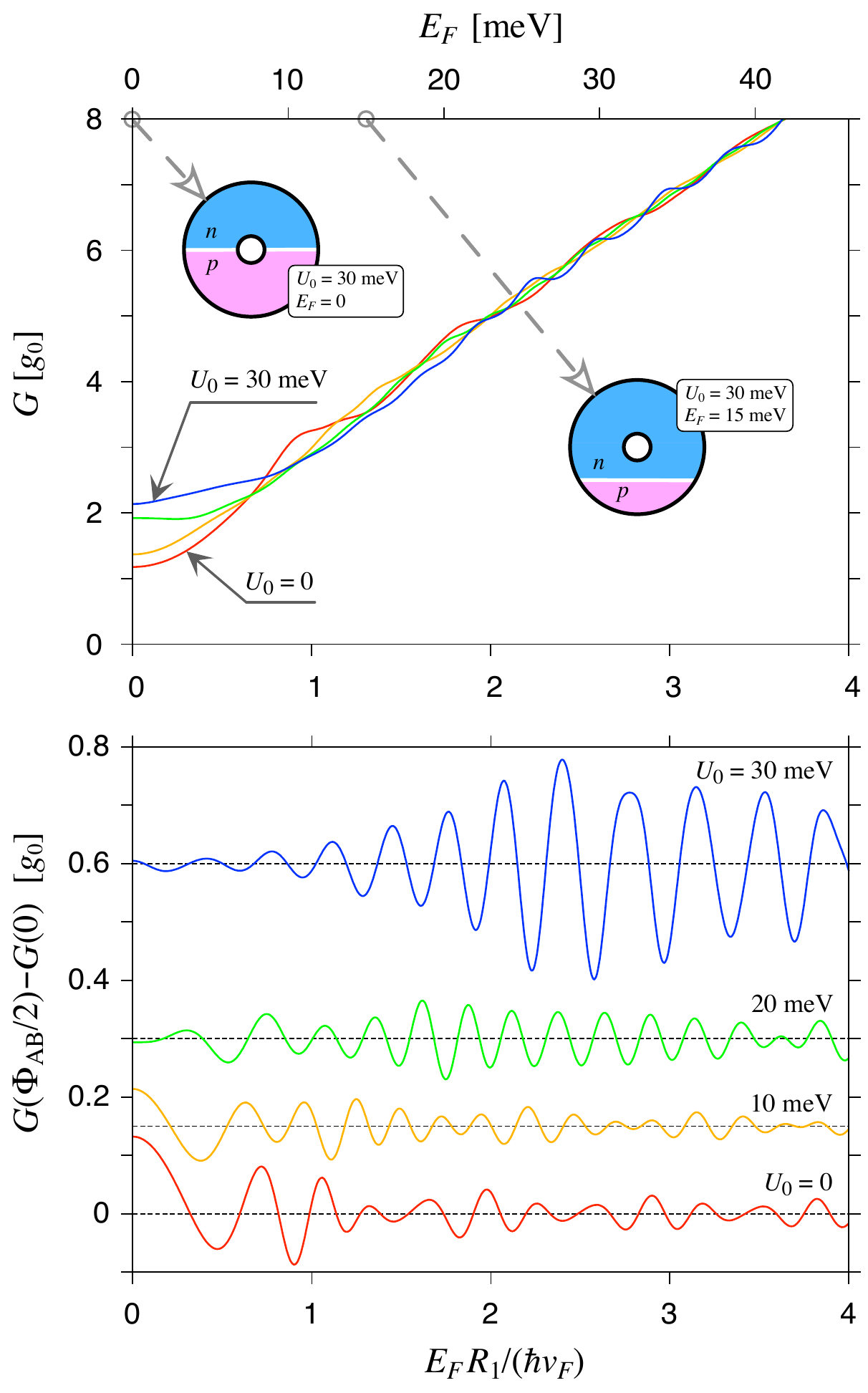}
\caption{ \label{delgr5vxx}
  Conductance for $\Phi_i=0$ (top) and the oscillation magnitude (bottom)
  displayed as functions of the Fermi energy for the disk radii
  $R_1=R_2/5 = 50\,$nm and the electrostatic potential amplitude
  [see Eq.\ (\ref{urphi})] varied from $U_0=0$ to $30\,$meV with the
  steps of $10\,$meV. 
  Top: Two insets show the positions of a~p-n interface in the disk area
  for $U_0=20\,$meV and the two different values of $E_F$. 
  Bottom:
  The datasets for $U_0>0$ (solid lines) are subjected to vertical shifts;
  black dashed lines show the actual $\Delta{}G\!=\!0$.
}
\end{figure}

So far, the discussion was limited to the case of a~perfect cylindrical
symmetry, allowing us to calculate the transmission probabilities $T_j$
[see Eq.\ (\ref{tjphi})]
analytically by solving the scattering problem separately for each
($j$-th) angular-momentum mode. 
In real system, several factors may break the cylindrical symmetry,
resulting in the mode mixing.
In particular, both the spatial corrugations of a~graphene sheet and
charge-donating impurities placed in the substrate lead to the charge
density fluctuations (i.e., {\em p-n puddles\/}) \cite{Dea10,Yua09,Sam16,Jay20}.
For best existing devices, carrier density fluctuations are
$\delta{}n<10^{11}\,$cm$^2$ near the neutrality point, corresponding
to the electrostatic potential fluctuation of the order of
$\delta{}U\sim{}10\,$meV. 

Here we test numerically, how robust are the effects which we describe in
Sec.\ \ref{resul} against the cylindrical symmetry breaking.
For this purpose, the electrostatic potential energy in the Hamiltonian
(\ref{hamdisk}) is replaced by \cite{Sus20}
\begin{equation}
  \label{urphi}
  U(r,\varphi) = -\frac{U_0r}{R_2}\sin\varphi, \ \ \ \ \ R_1<r<R_2. 
\end{equation}
In the leads, $r<R_1$ or $r>R_2$, we set $U(r,\varphi) =U_\infty$ again.
The potential amplitude (without loss of generality, we suppose
$U_0\geqslant{}0$) defines the Fermi energy range,
$-U_0<E_F<U_0$, for which a~p-n interface is present in the disk area
(see Fig.\ \ref{delgr5vxx}). A~special case of $U_0=0$ restores
the uniform-doping case considered in Sec.\ \ref{resul}. 

Regardless the value of $U_0$, angular-momentum eigenfunctions of the form
given by Eqs.\ (\ref{psirphi}), (\ref{chij1}) and (\ref{chij2}), still
represent the correct solutions in the leads.
Therefore, the numerical mode matching can be performed in the
angular-momentum space, employing the transfer matrix approach 
presented with details in Ref.\ \cite{Sus20}. 
Since the Fermi wavenumber $k_F=|E_F-U(r,\varphi)|/(\hbar{}v_F)$
is now position-dependent, the numerical results presented
in Fig.\ \ref{delgr5vxx} are parametrized by $E_F$ and $U_0$.
In order to specify these quantities in the physical units, we fixed
the disk dimensions at $R_1=50\,$nm and $R_2=5R_1=250\,$nm \cite{hbvfoo}.
However, it is worth to stress that the transport characteristics are
determined by the dimensionless parameters, $E_FR_1/(\hbar{}v_F)$
(also displayed in Fig.\ \ref{delgr5vxx}), 
$U_0R_1/(\hbar{}v_F)$, and the radii ratio $R_2/R_1$, and therefore
remain invariant upon the scaling $R_{1(2)}\rightarrow{}\lambda{}R_{1(2)}$,
$E_F\rightarrow{}E_F/\lambda$, and $U_0\rightarrow{}U_0/\lambda$, with 
a~real $\lambda>0$. 

If the system is close to the charge-neutrality point, namely for
$|E_F|<U_0R_1/R_2$, the conductance is noticeably enhanced with growing $U_0$
(see top panel in Fig.\ \ref{delgr5vxx}), as the propagation through heavily
p-doped and n-doped areas supplements the transport via evanescent waves.
(We further notice that the largest consider $U_0=30\,$meV corresponds to
$U_0R_1/(\hbar{}v_F)\approx{}2.6\lesssim{}R_2/R_1$, and thus the system,
at zero field, can be regarded as being in the crossover range between
the pseudodiffusive and the ballistic charge transport regimes
\cite{uu0foo}.)
For higher $|E_F|$, the effect of $U_0$ becomes negligible, and 
$G\approx{}2g_0\langle{}k_F\rangle{}R_1$ with
$\langle{}k_F\rangle{}=|E_F|/(\hbar{}v_F)$ being the average Fermi wavenumber on
the inner disk edge ($r=R_1$). 

The magnetoconductance oscillations magnitude (see bottom panel in Fig.\
\ref{delgr5vxx}) are diminished for $|E_F|<U_0R_1/R_2$ with growing $U_0$.
This observation can be rationalized by taking into account that at zero
magnetic field main currents flow along the $\varphi\approx\pm{}\pi/2$
directions (i.e., towards the regions of extreme doping), for which
the magnetic phases associated with the vector potential given by
Eq.\ (\ref{vecsol}) vanish.
In contrast, for the unipolar doping ($|E_F|>U_0$) the oscillations
are only weakly affected by growing $U_0$, and the magnitudes of
$\Delta{}G>0.1\,g_0$ appear for wide range of the doping.

\section{Conclusions}
\label{conclu}
We have demonstrated, performing the numerical analysis of the exact
formula for transmission probability for electron with a~given angular
momentum tunneling through the Corbino disk in graphene, 
that the conductance (as a~function of magnetic flux piercing the disk)
shows periodic oscillations of the Aharonov-Bohm kind.
Unlike for a~uniform magnetic field considered in Refs.\ \cite{Ryc10,Kat10},
when similar oscillations appear at discrete Landau levels only,
the disk in a~solenoid magnetic potential shows the oscillations for any
Fermi energy except from a~discrete energy set, defined by the disk radii
($R_1$ and $R_2$), the Fermi velocity in graphene ($v_F$) and the Planck
constant ($\hbar$), for which the conductance is flux-independent.  

Most remarkably, away from the charge-neutrality point the conductance
oscillations may show a~significant magnitude ($\Delta{}G>0.1\,g_0$,
with $g_0=4e^2/h$) starting from moderate radii ratios $R_2/R_1\gtrsim{}2$,
being comparable to the actual experimental values,
see Refs.\ \cite{Pet14,Kum18}. 
At the charge neutrality point, the oscillation magnitude grows with
the radii ratio, approaching $\Delta{}G\approx{}g_0$ for $R_2\gg{}R_1$. 

Also, we find out that the conductance oscillations are well-pronounced
in the presence of a~position-dependent electrostatic potential that breaks
the cylindrical symmetry and introduces the mode mixing.
Some suppression of the effect is predicted for ambipolar dopings (i.e.,
with a~p-n junction in the disk area), but the oscillations are restored
away from the charge neutrality point (for unipolar dopings).

\section*{Acknowledgments}
The work was supported by the National Science Centre of Poland (NCN)
via Grant No.\ 2014/14/E/ST3/00256.



\begin{thebibliography}{}
\bibitem{magfoo}
More generally, for a~centrosymmetric field ${\bf B}=B(r)\hat{e}_z$,
the fraction $\phi{}={\Phi_d/\Phi_0}$ in Eq.\ (\ref{tjphd})
is replaced by 
$$
  \phi\equiv\frac{\Phi_i}{\Phi_{\rm AB}}+\frac{4\pi}{\Phi_0}\int_{R_1}^{R_2}
  \frac{dr}{r}\int_{R_1}^r{dr'}r'B(r),
$$
where $\Phi_i$ is flux through the inner ring ($r<R_1$). In particular,
if $B(r)=0$ for $r>R_1$, the Aharonov-Bohm-like oscillations occur when
varying $\Phi_i$.
\bibitem{hbvfoo}
We took the nearest-neighbor hopping integral in graphene $t=2.7\,$eV
and the lattice parameter $a=0.246\,$nm, leading to
$\hbar{}v_F=\sqrt{3}\,ta/2=0.575\,$eV$\cdot$nm.
\end{thebibliography}
\end{document}